\documentclass[epj]{svjour}[24.12.01]
\usepackage{epsfig}
\usepackage{amssymb}
\usepackage{dcolumn}
\usepackage{amsmath}
\usepackage{tabularx}
\usepackage{units}
\usepackage{changebar}

\begin{document}
\def\etal{{\it et al.}}
\newcommand{\al}{\bar{\alpha}}
\newcommand{\be}{\bar{\beta}}
\newcommand*\prim{\,^\prime}
\newcommand*\thom{\frac{e^2}{m\,}\,}
\newcommand*{\textfrac}[2]{\ensuremath{\kern.1em
\raise.5ex\hbox{\the\scriptfont0 #1}\kern-.1em
/\kern-.15em\lower.25ex\hbox{\the\scriptfont0 #2}}}

\title{Quasi-free $\pi^0$ Photoproduction from
  the Bound Nucleon}
\author{K. Kossert\inst{1}\!$^,$\thanks{Part of the Doctoral  Thesis}$^,$
\!\!\thanks{\textit{Present address}:
Physikalisch-Technische Bundesanstalt,
Bundesallee 100, D-38116 Braunschweig} \and
M. Camen\inst{1}\!$^{,\rm a}$ \and
F. Wissmann\inst{1}\!$^{,\rm b}$
J. Ahrens\inst{2} \and
J.R.M. Annand\inst{3}
H.-J. Arends\inst{2} \and
R. Beck\inst{2} \and
G. Caselotti\inst{2} \and
P. Grabmayr\inst{4} \and
O. Jahn\inst{2} \and
P. Jennewein\inst{2} \and
M.I. Levchuk\inst{5} \and
A.I. L'vov\inst{6}   \and
J.C. McGeorge\inst{3} \and
A. Natter\inst{4} \and
V. Olmos de Le\'on\inst{2} \and
V.A. Petrun'kin\inst{6} \and
G. Rosner\inst{3} \and
M. Schumacher\inst{1}$^,$\thanks{e-mail: 
schumacher@physik2.uni-goettingen.de} \and
B. Seitz\inst{1}$^,$\thanks{\textit{Present
  address}: II. Physikalisches Institut der Universit\"at Gie\ss en, Germany} 
\and
F. Smend\inst{1}
A. Thomas\inst{2} \and
W. Weihofen\inst{1} \and
F. Zapadtka\inst{1}
}
\institute{II\@. Physikalisches Institut, Universit\"at G\"ottingen, D-37073
G\"ottingen, Germany \and
Institut f\"ur Kernphysik, Universit\"at Mainz,
D-55099 Mainz, Germany \and
Department of Physics and Astronomy, University of Glasgow,
Glasgow G12 8QQ, UK \and
Physikalisches Institut, Universit\"at T\"ubingen,
D-72076 T\"ubingen, Germany \and
B.I. Stepanov Institute of Physics, Belarussian Academy of
Sciences, 220072 Minsk, Belarus \and
P.N. Lebedev Physical Institute, 119991 Moscow, Russia
}
\date{Received: date / Revised version: date}

\abstract{Differential cross-sections for quasi-free $\pi^0$
photoproduction
from the proton and neutron bound in the deuteron have
  been measured for $E_\gamma = 200 - 400$ MeV at 
$\theta_\gamma^{\rm lab} =136.2^\circ$ using 
 the Glasgow photon tagger at MAMI,  the Mainz
  \unit[48]{cm} $\O$ $\times$ \unit[64]{cm} NaI(Tl) photon detector
  and the G\"ottingen SENECA recoil detector. For the proton
measurements made with both liquid deuterium and liquid hydrogen
targets allow direct comparison of ``free'' $\pi^0$ photoproduction
  cross-sections as extracted from the bound proton data with
  experimental
free
cross sections which are found to be in
  reasonable agreement below 320 MeV. At higher energies the ``free''
cross sections extracted from quasifree data are significantly smaller
than the experimental free cross sections and  theoretical predictions
based on multipole analysis. For the first time, ``free'' neutron
cross section have been extracted in the $\Delta$-region. They are also
in agreement with the predictions from multipole analysis up to 320
MeV and significantly smaller at higher photon energies.  
}

\maketitle
%\sloppy
%frist

%%%%%%%%%%%%%%%%%%%%%%%%%%%%%%%%%%%%%%%%%%%%%%%%%%%%%%%%%%%%%%%%%%%%%%
%%%%%%%%%%%%%%%%%%%%%%%%%%%%%%%%%%%%%%%%%%%%%%%%%%%%%%%%%%%%%%%%%%%%%%
%%%
%%%   M  A  I  N     T  E  X  T
%%%
%%%%%%%%%%%%%%%%%%%%%%%%%%%%%%%%%%%%%%%%%%%%%%%%%%%%%%%%%%%%%%%%%%%%%%
%%%%%%%%%%%%%%%%%%%%%%%%%%%%%%%%%%%%%%%%%%%%%%%%%%%%%%%%%%%%%%%%%%%%%%
\section{Introduction}

Single-pion photoproduction has been a subject of extensive experimental and
theoretical investigation for many decades. This reaction is one of 
the main sources of
information on  nucleon structure. It allows investigation of 
resonance excitations of the nucleon, especially the $\Delta(1232)$ 
excitation, and their photo-decay amplitudes. The
pion photoproduction amplitude is used as an input when calculating pion
photoproduction from  heavier nuclei and in the dispersion analysis of nucleon
Compton scattering. This reaction also serves as a test of
our understanding of the  chiral pion-nucleon dynamics. A well-known example
is the demonstration  that chiral perturbation theory accurately  
describes the 
very precise data on the $S$-wave multipole $E_{0+}$ and the $P$-wave 
amplitudes of $\pi^0$ photoproduction on the nucleon in the threshold region 
\cite{schmidt01,bernard96}.

Experimental investigation of pion photoproduction on the nucleon 
has a long history.
More than seventeen thousand data points form the modern data base of pion 
photoproduction at energies up to 2 GeV \cite{database}. Almost all of
them are data on the charged channels on both nucleons and 
on the $\gamma p\to \pi^0 p$
channel. The contribution  of the $\gamma n\to \pi^0 n$ reaction to
this base amounts to  only 120 data points at energies up to 905 MeV obtained 
in 1970's at the Frascati synchrotron
\cite{bacci72} and at the Tokyo synchrotron \cite{hemmi73,ando77}.
However,  no direct measurements of quasi-free  $\pi^0$
photoproduction on the neutron using a  deuterim target have been
carried out. Instead, the ratio 
$R=d\sigma (\gamma d\to \pi^0 np_s)/d\sigma (\gamma d\to \pi^0 pn_s)$
was measured where $p_s$ and $n_s$ are the spectator proton and
neutron, respectively,
and used to obtain the free-neutron cross-section based
on the assumption that $R$ is a good
approximation also for the ratio of the free cross-sections 
$d\sigma (\gamma n\to\pi^0 n)/d\sigma (\gamma p\to \pi^0 p)$. 
Though the method may be reasonable in principle, the data points
obtained from those works are of very limited  precision.

Following a proposal made in Ref.~\cite{LLP94} 
(see also Ref.~\cite{wissmann98}) an experiment has been carried out at
the 855 MeV microtron MAMI-B, where the primary goal  was to 
measure  differential 
cross-sections for  neutron  Compton scattering in  quasi-free 
kinematics using a deuterium target \cite{kossert01,camen02,kossert03}.
For that experiment the incident photon energy was chosen to be 
in the region from 200 MeV to 400 MeV. For the Compton scattering
experiment the   
produced $\pi^0$ mesons are  a source of
background photons due to their 2 $\gamma$ decay,
which has to be eliminated in the analysis. 
In the present paper
these events were used to obtain differential cross sections for
$\pi^0$ photoproduction from the proton and neutron bound in the deuteron.
By replacing the deuterium target
by a  hydrogen target    $\pi^0$  photoproduction
from the free proton was also measured under the same kinematic conditions.
It should be noted that the
present data are the first  where the quasi-free events  were
identified through a coincidence between one of the $\pi^0$
decay photons and the recoiling nucleon.

\section{Experiment}

\begin{figure}
%  \centering
%  \includegraphics[width=0.9\columnwidth]{figures/fig1.eps}
 \centerline{\epsfxsize=0.9\columnwidth\epsfbox{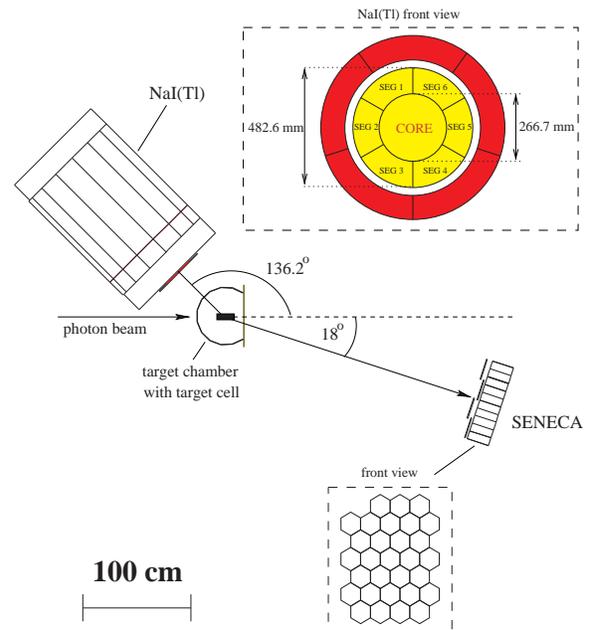}}
  \caption{\label{seneca_setup}The experimental setup used to
    measure quasi-free Compton scattering from the bound neutron and
    proton.  The scattered photons were detected with the large volume
    NaI(Tl) detector, the recoiling neutrons and protons with the
    SENECA detector system. Liquid deuterium and liquid hydrogen were used
    as target materials.  The target cells are mounted in a scattering
    chamber having a  Kapton window downstream the photon beam to
    reduce the energy loss of the protons on their way to SENECA.}
\end{figure}

The experimental arrangement installed at the Glasgow photon tagger at
MAMI \cite{anthony91} and outlined in Fig.~\ref{seneca_setup}
has been described  previously \cite{kossert01,camen02,kossert03}. 
This allows us to give here only a short description.
The large Mainz \unit[48]{cm}
$\O$ $\times$ \unit[64]{cm} NaI(Tl) detector \cite{wiss94,wissmann99} was
positioned at a scattering angle of 
$\theta^{\rm lab}_\gamma = 136.2^\circ$ (nominally $135^\circ$). The
energy resolution of this detector is \unit[1.5]{\%} in the $\Delta$
energy region and its detection efficiency \unit[100]{\%}.  The recoil
nucleons were detected with the G\"ottingen SENECA  detector
\cite{seneca} positioned at an average emission angle of $\theta_N =
-18^\circ$, thus covering the angular range corresponding to  quasi-free
kinematics.  This angle was optimized for Compton scattering but sufficiently
covers also the range required for  $\pi^0$ photoproduction where the
average emission angle is $\theta_N = -15^\circ$. Effects of the
variation of the emission angle are precisely taken into account 
by  computer simulation.
A target to recoil detector distance of \unit[250]{cm}
was  chosen as a compromise between the  energy resolution for the
time-of-flight measurement, $\Delta E_{n}/E_{n} \approx
\unit[10]{\%}$, and the geometrical acceptance
$\Delta\Omega_{n}\approx \unit[90]{msr}$.  As target a \unit[5]{cm}
$\O$ $\times$ \unit[15]{cm} Kapton cell filled with liquid deuterium
was used. By filling the same cell with liquid hydrogen it was possible
to investigate quasi-free and free $\pi^0$ photoproduction on the proton 
under  identical kinematic conditions.

SENECA consists of 30 hexagonally shaped cells 
filled with NE213 liquid scintillator  
(\unit[15]{cm} in diameter and \unit[20]{cm} in length)
mounted in a honeycomb structure.
Veto-detectors in
front of SENECA provided the possibility to identify charged
background particles and to discriminate between neutrons and protons.
This allowed  clean separation  between quasi-free Compton scattering
and $\pi^0$ photoproduction from the proton and neutron detected in
the same experiment.  The detection efficiency of the veto-detectors
for protons was implemented in the Monte Carlo program \cite{brun93}.
The detection
efficiency for neutrons was measured {\em in situ} by analysing 
events from the
$p(\gamma,\pi^+ n)$ reaction leading to $\epsilon_n = 18\%$
\cite{kossert03} on the average. The new data for   $\epsilon_n$    being
valid on a few percent level of precision are used 
to correct the results of computer simulations. For further details see
\cite{kossert03}. 
The momenta of the recoil nucleons were  measured using the
time-of-flight (TOF) technique with the NaI(Tl) detector providing the
start signal and the SE\-NE\-CA modules providing the stop signals.

Data were collected during \unit[238]{h} of beam time with a deuterium
target and  \unit[35]{h} with a  hydrogen target. The tagging
efficiency, being \unit[55]{\%}, was measured several times during the
runs by means of a Pb-glass detector moved into the direct photon
beam, and otherwise monitored by a P2 type ionization chamber
positioned at the end of the photon beam line.

\section{Data analysis}

Before analysing the data obtained with the deuterium  target the
corresponding analysis of data obtained with the hydrogen target was
carried out.  In this case the separation of events from Compton
scattering and $\pi^0$ photoproduction can  be achieved by the
NaI(Tl) detector alone, but the detection of the recoiling proton improves
the separation, especially for energies near the peak of the $\Delta$
resonance.  A typical spectrum is shown in panel {\bf a} of
Fig. 2. The data obtained for the free proton have been used to
optimize the analysis procedure for the bound nucleon.

\begin{figure}
 % \centering
 % \includegraphics[width=1.0\columnwidth]{figures/fig2.eps}
 \centerline{\epsfxsize=1.0\columnwidth\epsfbox{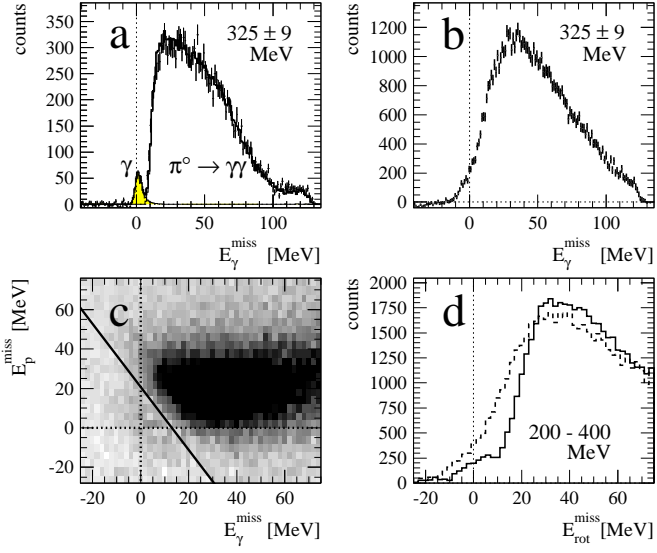}}
  \caption{\label{miss_2d} Panel {\bf a}:
    Number of  proton events obtained with a hydrogen target versus
    the missing photon energy $E^{\rm miss}_{\gamma}$. Panel {\bf b}: 
    The same as panel {\bf a} but
    for proton events obtained with a deuterium  target. 
    Panel {\bf c}: Scatter plot of
    proton events obtained with a deuterium  target. Abscissa and
    ordinate are the missing energies of the scattered photon and the
    recoil proton, respectively.  The thick solid line separates
    Compton events located in the vicinity of the origin 
    and $(\gamma,\pi^0)$ events shown as a dark
    range. For the further evaluation each point in the scatter plot
    was rotated (moved on a circle centered at  the origin), until the
    thick solid line was perpendicular to the abscissa.
    Panel {\bf d}:
    Projection of proton events obtained with a deuteron target after
    rotation (solid line) and before rotation (broken line).}
\end{figure}
\begin{figure}
%  \centering
%  \includegraphics[width=1.0\columnwidth]{figures/fig3.eps}
 \centerline{\epsfxsize=1.0\columnwidth\epsfbox{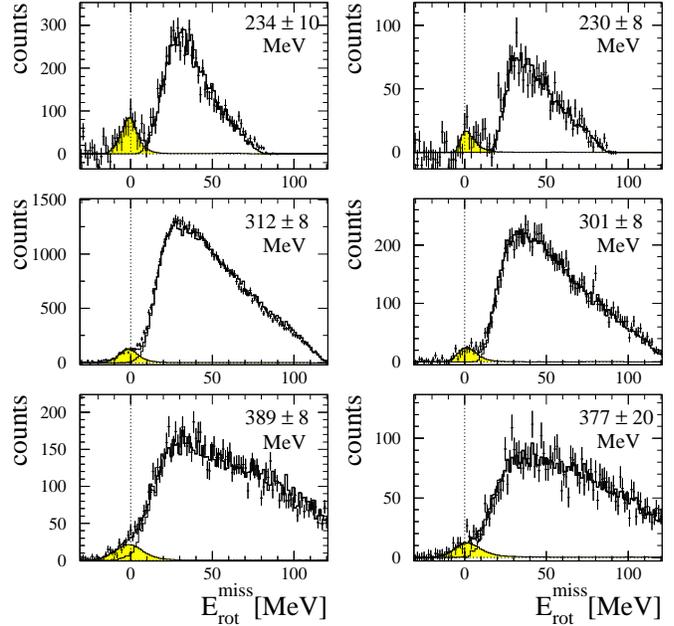}}
  \caption{\label{miss_1d}Typical spectra of events obtained with
    the deuteron target shown for the recoil proton (left panels) and
    recoil neutrons (right panels). The data from the two dimensional
    plot have been projected on to the abscissa after the rotation 
    described in the caption of Fig.2 and in the text. 
    The solid curves are the results of a Monte
    Carlo simulation scaled to the Compton events and the
    $(\gamma,\pi^0)$ events, respectively.}
\end{figure}

For the separation of events from Compton scattering and $\pi^0$
photoproduction it is convenient to use two-dimensional scatter plots
of events with the missing nucleon  energy 
$E^{\rm miss}_N = E^{\rm calc}_N  - E^{\rm SEN}_N$ 
and the missing photon energy  
$E^{\rm miss}_\gamma = E^{\rm calc}_\gamma -E^{\rm NaI}_\gamma$ 
as the parameters, where $E^{\rm SEN}_N$ and $E^{\rm NaI}_\gamma$
denote measured energies and $E^{\rm calc}_N$ and $E^{\rm calc}_\gamma$
the corresponding calculated energies. The calculations are carried out
using the tagged photon energy and the detected nucleon angle or  scattered
photon angle and 
assuming the kinematics
of Compton scattering by the proton  in case of a hydrogen target or the
kinematics of Compton scattering in the center of the quasi-free peak
\cite{LLP94}
in case of a deuteron target. As an example  panel {\bf c} of
Fig.~\ref{miss_2d} shows the scatter plot of proton events obtained with
\unit[200]{MeV} $-$ \unit[400]{MeV}  photons  incident on a
deuterium target.  Two
separate regions containing events are visible.  The Compton events
are located in a narrow zone around the origin ($E^{\rm miss}_\gamma =
0, E^{\rm miss}_p = 0$), the $(\gamma,\pi^0)$ events in the dark range
at larger missing energies.  For the further evaluation each point in
the panel was rotated (moved on a circle centered in the origin) until
the thick solid line became perpendicular to the abscissa. This has
the advantage that projections of the data on the new abscissa 
-- denoted by $E^{\rm miss}_{\rm rot}$ --
can be used
for further analysis without  loss in the quality of the
separation between  the two types of events.

The benefits of this procedure are illustrated in 
Fig. 2.  Panel {\bf a} shows numbers of proton events 
from a proton target versus the measured
missing energy of the scattered photon, given for a narrow 
energy interval close to
the maximum of the $\Delta$ resonance. In this case we find  very
good separation between  the two types of events as in previous
experiments carried out with proton targets.  The good separation
disappears when proton events of the same type are taken from a
deuteron target. These data are shown in panel {\bf b} where it can
clearly be seen that the effects of binding destroy the separation of
the two types of events which was visible in panel {\bf a}. The separation
can be partly restored when the rotation
procedure is applied as is shown in panel {\bf d}. This panel contains as a
solid line the same data as the scatter plot of events 
shown in panel {\bf c} but projected
on to the abscissa after rotation. For comparison the broken line shows
the same data before rotation. The comparison of these two lines
clearly demonstrates that the rotation procedure  improves
the separation of the two types of events. 

Fig.~\ref{miss_1d} shows typical spectra  obtained with the
deuterium target. The left panels contain proton events, the right panels
neutron events. The different numbers of events on the two sides are
due to the difference between the SENECA detection efficiency 
for neutrons ($\approx$ \unit[18]{\%})        
and protons ($\approx$ \unit[99] {\%}).
There is a reasonable separation of the two types of events in the
whole energy range. For the final separation and for the determination
of the numbers of photopion  events a complete Monte
Carlo simulation has been carried out for the processes under
consideration.  The results of these
simulations shown by solid curves were scaled to the Compton and
$(\gamma,\pi^0)$ data, thus leading to the grey areas in case of the
Compton scattering events and to the white areas in case of the
$\pi^0$ events.

The number of $(\gamma,\pi^0)$ events which is the number of events
given by the adjusted curves, corresponds to the integral of the
triple differential cross-section in the region of the quasi-free
peak. The following relation has been used to determine the final
triple differential cross-section in the center of the nucleon
quasi-free peak (NQFP):
\begin{equation}
\left( \frac{d^3\sigma}{d\Omega_{\pi^0}d\Omega_NdE_N}\right)_{\rm NQFP}
=\frac{N_{\pi^0 N}}{N_{\gamma}N_T \,\epsilon_N R^{\gamma\pi^0}_{\rm NQFP}},
\end{equation}
where $N_{\pi^0 N}$ is the number of coincident $\pi^0$-nucleon
events as extracted from the missing energy spectra, $N_\gamma$ is
the number of incident photons, $N_T$ is the number of target nuclei, 
$\epsilon_N$ is the nucleon detection efficiency and 
$R^{\gamma\pi^0}_{\rm NQFP}$ is a factor obtained by Monte Carlo
simulation which relates the number $(\gamma,\pi^0)$ events integrated
over the distribution of events to the triple differential cross-section in the center of the nucleon quasi-free peak.

\section{Theory}

\begin{figure}
%  \centering
%  \includegraphics[width=1.\columnwidth]{compton_graphs2.eps}
  \centerline{\epsfxsize=1.0\columnwidth\epsfbox{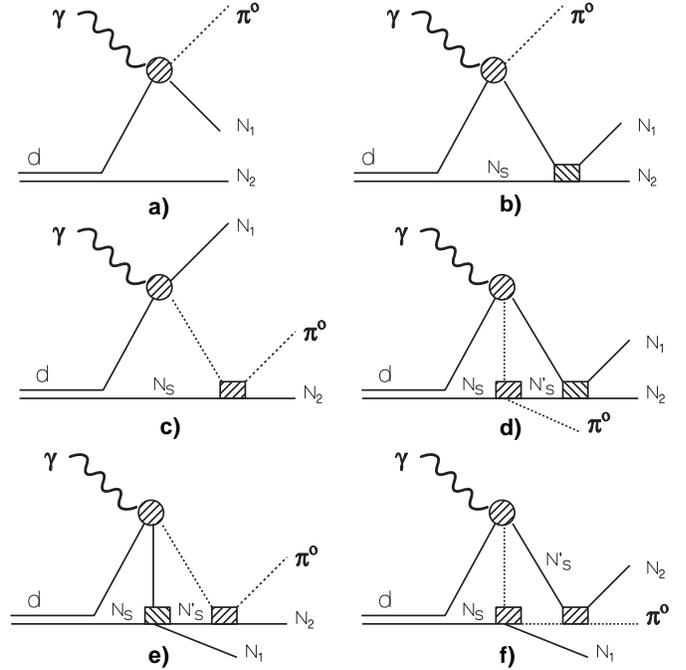}}
\vspace{1cm}
  \caption{\label{graphs}
Graphs contributing to the reaction $\gamma d\to \pi^0 np$. The
set of graphs where  $N_1 \leftrightarrow N_2$ is not shown in the
figure. }
\end{figure}

In our analysis of the data we  used the  theoretical
model proposed in Refs.~\cite{laget78,laget81} and developed
subsequently in Refs.~\cite{LLP96,LSW00,LSW00_2} (see also
Refs.~\cite{schmidt96,darwish02}). The model is based on 
the diagrams   relevant for  the reaction on 
the kinematic conditions under consideration. The main graphs
contributing to the reaction amplitude in the quasi-free region
are  displayed in Fig.~\ref{graphs}. Graph
\ref{graphs}a) describes quasi-free photoproduction from the
nucleon $N_1$. It is expected to be dominant when the momentum 
of the  nucleon
$N_2$ is sufficiently  small ($\lesssim \sqrt{mE_b}$ where
$E_b=2.2246$ MeV is the deuteron binding energy and $m$ the nucleon
mass). 
This corresponds to  the
so-called nucleon quasi-free peak (NQFP) region. The non-interacting
nucleon $N_2$ in this graph is often referred to as a spectator.

The final nucleons can interact with  each other through the
mechanism displayed in graph \ref{graphs}b), describing the so-called
final state interaction (FSI). This graph was found to be of great
importance for  many processes involving  deuteron disintegration,
including the reaction $\gamma d\to \pi NN$ (see
Refs.~\cite{laget78,laget81,LLP96,LSW00,LSW00_2,darwish02}). The
big effect of FSI is mainly caused by $NN$-interaction in the
$s$-wave at small relative momenta ($\lesssim 200$
MeV/c) of the $NN$-pair. Such  small
relative momentum is provided under the kinematic conditions of small
incident energies and/or forward angles of a third particle in the
final state, which is the pion in our case. Of the two $s$-wave final-states, 
$^1S_0$ and $^3S_1$, the
repulsive isosinglet $np$-wave $^3S_1$ proved to be the more important
one,
leading to a  decrease of the cross-section due to FSI.
Experiments on semi inclusive $\pi^0$ photoproduction in the
reaction $d(\gamma,\pi^0)np$ \cite{krusche99,siodl01} clearly
confirm these properties of FSI.  Since our experiment covers the
kinematic region where the relative momentum of the $np$-pair
ranges from 140 to 270 MeV/c we expect that FSI  gives a noticeable
contribution at the beginning of  the energy interval and decreases in
importance with increasing  photon energy.

The pion (neutral or charged) produced in  the $\gamma N\to \pi
N$-vertex can be scattered  by the spectator nucleon as  displayed in
graph \ref{graphs}c).  
Although the  detailed investigation of the reaction 
$\gamma d\to \pi^-pp$ performed in  
Refs.~\cite{laget78,laget81} has shown that there
exists a kinematic region where the diagram  \ref{graphs}c) can give a
noticeable contribution to the amplitude  in the $\Delta$ region,
usually the effect of $\pi N$-rescattering is smaller than that 
of $NN$-rescattering. 
In our previous paper \cite{LLP96} we found the contribution of graph
\ref{graphs}c) in the NQFP region to be of minor importance at energies from
250 to 400 MeV.

It is known that two-loop diagrams  can be important for the description of
the reaction $\gamma d\to \pi NN$ under  certain kinematic conditions.
For instance, the graph \ref{graphs}d) proved to give a big contribution to the
inclusive process $d(\gamma,\pi^0)np$ in the threshold region (see
Ref.~\cite{LSW00}).
This result can be easily understood if one takes into account that the
threshold electric dipole amplitude $E_{0+}$ of charged pion photoproduction
which is contained  in the upper block of this graph,
is about 30 times larger in  absolute numbers  than those for 
neutral channels.  
With increasing  photon energy this effect is expected to decrease
in  importance. Nevertheless, we will take it into account.

A significant modification of the reaction amplitude through s-wave
interaction of the NN-pair  may not only occur in the final
state but also in the intermediate state by
the mechanism displayed in  graph  \ref{graphs}e). This was 
previously demonstrated in Refs.~\cite{laget78,laget81} in the
analysis of the reaction $\gamma d\to \pi^- pp$ where it was shown
that of the two possible $s$-wave interactions, $^1S_0$ and $^3S_1$, 
the isosinglet $^3S_1$-state is  dominant. We have observed the
same result in our model. 
It should be noted that, for example in the pQFP region, almost the total
contribution from the graph \ref{graphs}e) stems from the configuration where
the nucleon $N_1$ is the neutron, i.e. this nucleon has a small momentum. 
In such a situation, the kinematics allows both the deuteron wave function
(DWF) and $np$-scattering amplitude to act  simultaneously in the low momentum
regime where they are strongly enhanced.

There is  one further  two-loop contribution due to $\pi N$-rescattering  
both in the intermediate and final states and this  is displayed in  
diagram \ref{graphs}f). 
Since the $\pi N$-scattering amplitude is smaller than the 
$NN$-scattering amplitude we expect this diagram to be less important
in comparison with
e.g. diagram  \ref{graphs}d). Below we will see that the latter gives 
only a few
percent contribution to the differential cross-section. Therefore, the diagram
\ref{graphs}f) can  safely be disregarded.
The smallness of the contribution of  this diagram  to the amplitude
of the reaction
$\gamma d\to \pi^- pp$ in the $\Delta$ region was also mentioned
in Refs.~\cite{laget78,laget81}.

All details of the calculations in the extended model, i.e.
including the graphs c) to f) of Fig.4,  will be published 
elsewhere \cite{LSW03}. 
Here we mention only the following. The pion photoproduction amplitude was
taken in the on-shell form and calculated with the SAID
\cite{arndt02} and MAID \cite{drechsel99} multipole analyses. Below we will
use the most recent SAID SM02K and MAID DMT2001 solutions. 
The off-shell corrections  are expected to be small in the NQFP region 
so that the use of the on-shell parametrization for the amplitudes is quite
justified.

A model of the $NN$-interaction is needed to calculate the DWF 
and the $NN$-scattering amplitude when evaluating the diagrams. We
checked three versions of the Bonn OBEPR model \cite{OBEPR1,OBEPR2},
the CD-Bonn potential \cite{CD-Bonn}, and a separable approximation
\cite{PEST1,PEST2} of the Paris potential \cite{Paris}.  Our
observation is that the results obtained with all the potentials
are practically the same so that in the following we will present our
results with the most recent of them, namely with the CD-Bonn model.

The $\pi N$-scattering amplitude was calculated in a meson-exchange model
\cite{hung94,hung01} constructed in the three-dimensional Bethe-Salpeter 
formulation.
To take into account the off-shell nature of the intermediate pion, we supply
its propagator by a dipole-like form factor,
$F_\pi(q_{on},q_{off})=(\Lambda^2_\pi + q^2_{on})^2/ (\Lambda^2_\pi + q^2_{off})^2$
with $ q_{on}$ ($q_{off}$) being the on-shell (off-shell) momentum of the
intermediate pion. Introducing the form factor ensures also the 
convergence of the 
integrals over the pion off-shell momentum ${\bf q}_{off}$ which emerge in the
evaluations of graphs \ref{graphs}c)-d). The value of the cut-off 
parameter $\Lambda_\pi$ is
usually treated as a free parameter which is adjusted to provide the best
description of the  reaction under consideration. It is not surprising 
that in the literature there
exists  a great variety of numerical values for $\Lambda_\pi$. 
We use two different choices for it. The first one is a very soft 
form-factor ($\Lambda_\pi=440$ MeV) which was used in
Refs.~\cite{kamalov99,kamalov01} to give the best fits to the  $\pi^0$   
photo and
electroproduction data in the threshold region as well as to the $\Delta(1232)$
resonance multipole $M^{(3/2)}_{1+}$ over a wide energy range. The
second one is a very
hard form-factor ($\Lambda_\pi=1720$ MeV) used in  Ref.~\cite{CD-Bonn} for the
construction of the CD-Bonn potential. 
We found, however, that  the variation of  $\Lambda_\pi$ from 1000  MeV  to
1720 MeV practically does not change the results. Moreover, for $\Lambda_\pi
>1000$ MeV one can safely put the form factor $F_\pi(q_{on},q_{off})$ to be
equal to 1. In actual calculations, the upper limit of the integrals
was taken to be $p^{max}=1000$ MeV/c but its replacement by   
$p^{max}=500$ MeV/c
changed  the cross-sections by less than 3\%.

\begin{figure}
%  \centering
%  \includegraphics[width=1.\columnwidth]{compton_graphs2.eps}
\centerline{\epsfxsize=1.\columnwidth\epsfbox{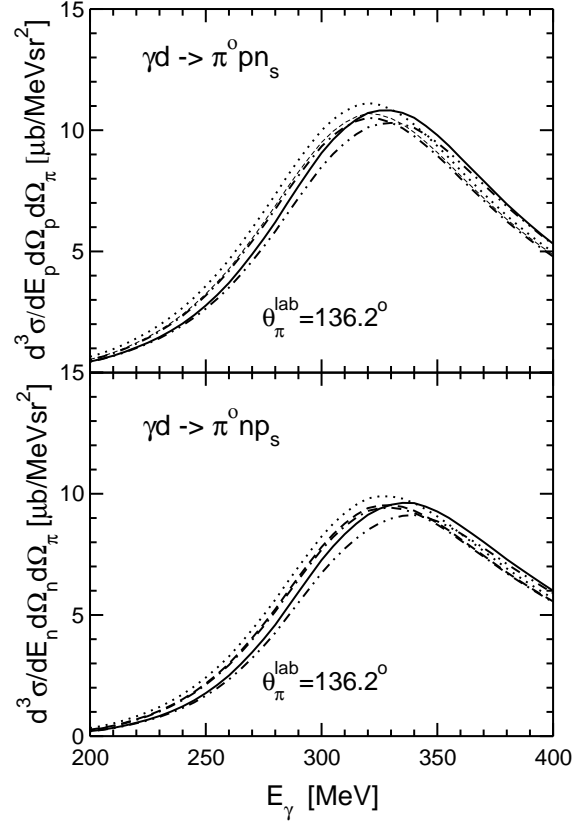}}
  \caption{\label{contrbs}
Contributions of the graphs in Fig.~ \ref{graphs} to the triple
differential cross-section of the reaction $\gamma d\to \pi^0 np$
in the CpQFP (upper figure) and the CnQFP (lower figure). Dotted curves: pole
proton (upper figure) and neutron (lower figure) diagrams (graph 4a). 
Successive 
addition of
graphs \ref{graphs}b), d), and e) gives the dashed, dash-dotted, and solid 
lines, respectively.
The contribution from graph \ref{graphs}c) is  small and
is not shown in the figure. All the curves are obtained for $\Lambda_\pi=440$
MeV. The total results for $\Lambda_\pi=1720$ MeV is shown in
dash-double-dotted curves.
}
\end{figure}
 
Contributions of separate diagrams to the triple differential 
cross-section of the reaction $\gamma d\to \pi^0 np$ 
are shown in Fig.~\ref{contrbs}.
The effect of FSI manifests itself in a noticeable reduction of the 
cross-section. The size of this reduction
ranges from 20\% at 200 MeV, and 5\% at 300 MeV, to 2\% at 400 MeV
in the center of the proton quasi-free peak (CpQFP). In the center
of the neutron quasi-free peak (CnQFP) one has the same numbers. It is
interesting to note that the numbers given above were  also 
obtained by us for the relative FSI
contribution in the case of the reaction $\gamma d\to \gamma' np$ 
\cite{kossert03}. 
The effect of $\pi N$-rescattering (graph \ref{graphs}c)) does not exceed 1\%
and  is not shown in  Fig.~\ref{contrbs}. The inclusion of graph \ref{graphs}d)
leads to a decrease of the cross-section by 6\% at 200 MeV, 2\% at 300 MeV, and
0.5\% at 400 MeV. These numbers are practically independent of the
$\Lambda_\pi$ value. A further reduction of the cross-section below 320 MeV is
due to the contribution of diagram \ref{graphs}e). The reduction  is 
quite visible, being about 15\%
from 200 to 260 MeV and reducing further to 8\% at 300 MeV.  These numbers
are obtained for $\Lambda_\pi=1720$ MeV. For $\Lambda_\pi=440$ MeV they are
12\% and 2\%, respectively. Above 320 MeV one observes some increase of the
cross-section after inclusion of graph \ref{graphs}e). The total contribution
of one-loop and two-loop diagrams is --30\%, --15\%, and +6\% at 200, 300, and
400 MeV, respectively, if one takes 1720 MeV for  $\Lambda_\pi$. 
With  $\Lambda_\pi=440$
one has almost the same numbers  at 200 and 400 MeV but at 300 MeV the 
decrease of the
cross-section is noticeably smaller being about --9\%. 

The measured triple differential cross-section,\\ $d^3\sigma /
d\Omega_{\pi}d\Omega_p dE_p $ (here $E_p$ and $\Omega_p$ are the
kinetic energy and solid angle of the proton), in the CpQFP can be
related to the differential cross-section of pion
photoproduction on the ``free'' nucleon, 
$d\sigma /
d\Omega_{\pi}$, via a spectator formula
\begin{eqnarray}
 \lefteqn { \frac{ d\sigma(\gamma p\rightarrow \pi^0 p) }
  { d\Omega_{\pi} } = }  \nonumber \\
 &&\frac{ (2\pi)^3 }{ u^2(0) }
  \frac{ E_\gamma |{\bf q}_\pi|^2 }
{ |{\bf p}_p|E^f_\gamma (\varepsilon_\pi q_\pi \cdot
p_p-\varepsilon_p\mu^2)}
  \frac{ d^3\sigma(\gamma d\rightarrow \pi^0 p n) }
  {d\Omega_{\pi}d\Omega_p dE_p },
  \label{spectator}
\end{eqnarray}
where $u(0)$ is the S-wave amplitude of the DWF at zero momentum 
(the D-wave component of this function does not
contribute at zero momentum), $q_\pi$ is the pion 4-mo\-men\-tum,
$\mu$ is the pion mass, $p_p$ is the proton 4-momentum.
$E^f_\gamma$ is the lab photon energy corresponding to free-pion 
photoproduction
\begin{equation}
  E^f_\gamma=\frac {(p_p+q_\pi)^2-m^2}{2m}=E_\gamma -E_b \left( 1 + \frac {E_\gamma - E_b/2}{m} \right).
  \label{energy}
\end{equation}
In the case of quasi-free pion photoproduction on the neutron, one has
a formula analogous to that of Eqs.~(\ref{spectator}) and
(\ref{energy}) 
but with the replacement
$p\leftrightarrow n$.

Equation~(\ref{spectator}) is valid for the pole diagram contribution
(Fig. 4a) only. Therefore, in order to make Eq.~(\ref{spectator}) valid for
practical applications
the r.h.s.\ of this equation has to be
multiplied by a factor
$f(E_\gamma,\theta_\gamma)=d^3\sigma_{pol}/d^3\sigma_{tot}$
(see analogous discussion in Ref.~\cite{kossert03}).
Here,
$d^3\sigma_{pol}$ stands for the contribution of the pole (proton or
neutron) diagram to the total differential cross-section
$d^3\sigma_{tot}$ for which all the diagrams a)-f) have been taken
into account.

\section{Discussion of the Results}

\begin{figure}
%  \centering
%  \includegraphics[width=1.0\columnwidth]{qf_pi0.eps}
\centerline{\epsfxsize=1.\columnwidth\epsfbox{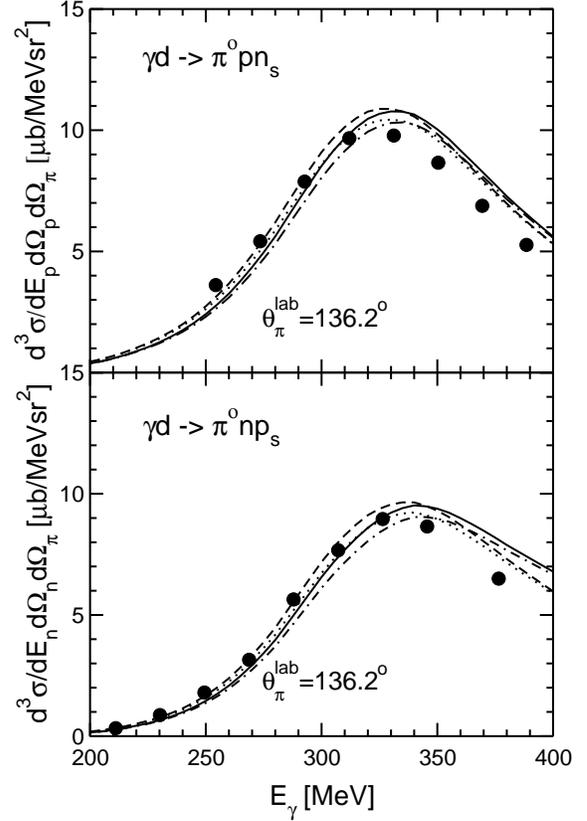}}
  \caption{\label{cs_qfree}
Triple differential cross-section of the reaction
    $d(\gamma, \pi^0 p)n$ (upper figure) and $d(\gamma, \pi^0
    n)p$ (lower figure) in the center of the NQFP at
    $\theta^{\rm lab}_\pi = 136.2^\circ$. Results for the MAID2001 solution
    are shown in dash-dotted (with $\Lambda_\pi=440$ MeV) and 
solid (at $\Lambda_\pi=1720$
    MeV) curves. Corresponding results for the SM02K solution 
are given in dashed and dotted curves.
}
\end{figure}

Results for the triple differential cross-section of the reactions
$d(\gamma, \pi^0 p)n$  and $d(\gamma, \pi^0n)p$ in CpNQFP and CnQFP
at $\theta^{\rm lab}_\pi = 136.2^\circ$ are given 
in Tables~\ref{table_qf_proton_pi0} and
\ref{table_qf_neutron_pi0}, respectively, and displayed 
in Fig.~\ref{cs_qfree}. 
Also in this figure we show theoretical predictions in the framework of the
model described above. The area  filled by the curves gives the size of
uncertainties of the theoretical model.
One can see good agreement between the experimental data and theoretical
predictions below 320 MeV. This agreement, however, vanishes  above the
$\Delta$-peak. 
At present we do not know the reasons responsible for the
disagreement.
For instance, the effect of $N\Delta$-interaction omitted in our theoretical
model was shown in Ref.~\cite{obukh03} to be of no importance in the 
$\Delta$-region at backward angles.  
Further theoretical efforts are needed to shed light on the above situation.

\begin{figure}
%  \centering
%  \includegraphics[width=1.0\columnwidth]{free_pi0.eps}
\centerline{\epsfxsize=1.\columnwidth\epsfbox{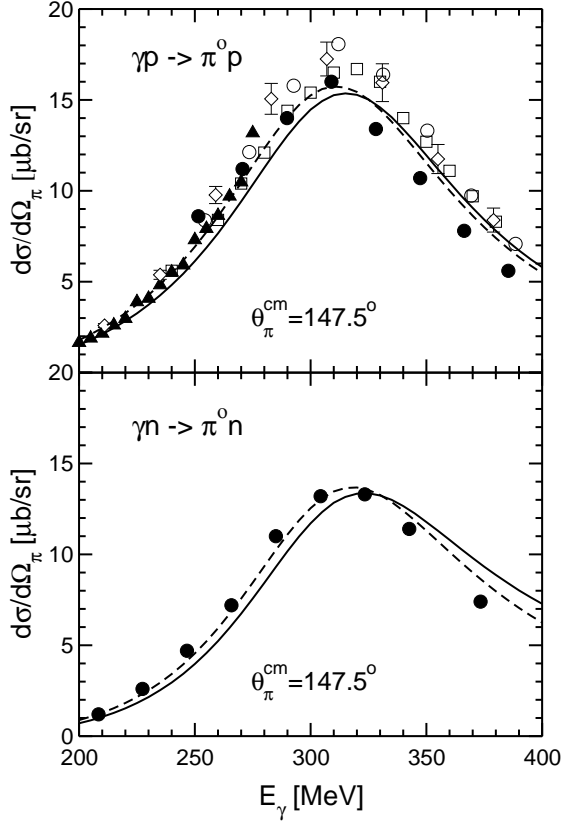}}
  \caption{\label{free_pn}
The CM differential cross-section of the free proton (upper figure)
    and free neutron (lower figure) at $\theta^{cm}_\pi =
    147.5^\circ$. The present free data measured with the
    hydrogen target are shown as open circles. Present ``free'' data 
extracted from
    cross-sections for the bound nucleon are shown as filled circles. 
    Also shown are the available data from hydrogen targets from  
Refs.~\cite{genzel74} ($\Box$), 
       \cite{fuchs96} ($\triangle$), and \cite{haerter96} ($\Diamond$)
measure with a  hydrogen target. The solid
       and dashed curves represent results obtained from the  MAID2001 
and SM02K solutions, respectively.} 
\end{figure}

Using Eq.~(\ref{spectator}) with the correction factor 
$f(E_\gamma,\theta_\gamma)$ included,
we can extract the free-nucleon cross-sections from
the corresponding quasi-free data. 
It should be noted that the extracted free values are practically 
independent of the choice
of the  multipole analysis of pion photoproduction so that the only model
dependence in these values stems from their sensitivity to the cut-off 
parameter $\Lambda_\pi$. 
After averaging over two sets of results for $\Lambda_\pi=440$ and 1720 MeV we
obtain the central numbers given in Tables~\ref{table_qffr_proton_pi0} and
\ref{table_qffr_neutron_pi0}  and displayed in Fig.~\ref{free_pn}
to which an uncertainty of about 4\% due to the conversion from
quasi-free to ``free''
according to Eq.~(\ref{spectator}) should be attributed. 
In  Fig.~\ref{free_pn} we also show the cross-sections measured with the
hydrogen target (see Table~\ref{table_free_proton_pi0}) and compare 
them with the data obtained in other experiments
\cite{genzel74,fuchs96,haerter96} and with the predictions of the multipole
analyses.  Reasonable agreement between all data sets is seen up to 
the $\Delta$-peak. 
All of them fairly well correspond to the SM02K solution and 
are  slightly above the
predictions of the MAID2001 solution. However, above 320
MeV our quasi-free data points for the proton lie significantly below
both all the free data and the multipole predictions. Of course, this has
to be expected from the unsatisfactory  description of the 
quasi-free data above 320 MeV mentioned above. The free-neutron 
cross-sections are  consistent
with the multipole predictions at all energies except for  377 MeV 
where the measured
value is noticeably smaller than the predicted one.

\begin{figure}
%  \centering
%  \includegraphics[width=1.0\columnwidth]{fig8.eps}
\centerline{\epsfxsize=1.\columnwidth\epsfbox{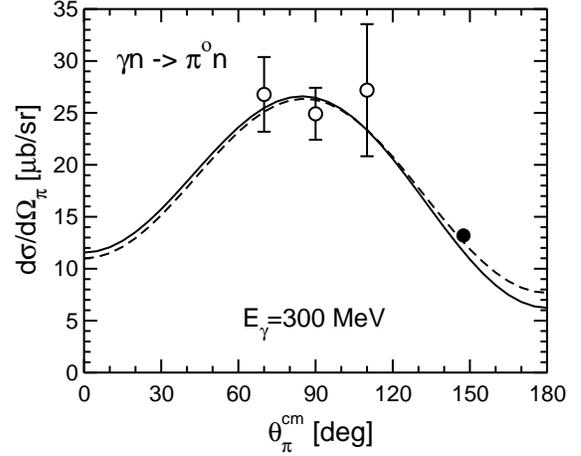}}
\vspace{-5cm}
  \caption{
\label{free300n}
The angular dependence of the CM differential cross-section of the 
reaction  $\gamma n\to\pi^0n$ at 300 MeV. The present data point is shown 
as the  filled circle. 
Data from Ref.~\cite{ando77} are shown as empty   circles. Solid
       and dashed curves represent results of the  MAID2001 and SM02K
       solutions, respectively.} 
\end{figure}

As has been mentioned  in the introduction, the experimental information on 
the  $\gamma n\to\pi^0n$ channel is very
sparse. Up to now there has been only one measurement of the differential
cross-section in the $\Delta$-region \cite{ando77} which covered the angular
region from $70^\circ$ to $130^\circ$.  In  Fig.~\ref{free300n} we show the
angular distribution of the cross-section at 300 MeV. One can see that the data
from  \cite{ando77} and from the present paper are consistent with
each  other in the sense that  they are in agreement with the 
predictions of the same multipole analyses. But a significant
improvement in the accuracy of the present experiment in comparison with that
of  Ref.~\cite{ando77} is evident.

\section{Conclusion}

The energy dependence of the triple differential cross-section for
$\pi^0$ photoproduction from the proton and neutron in deuterium
and the free photoproduction cross section from the proton have
been measured in the same kinematics at $\theta^{\rm lab}=136.2^\circ$. 
For the first time accurate
``free'' cross-sections have been extracted for the proton and the
neutron from the
quasi-free data in the $\Delta$ region, using an extended model for the
conversion developed in the present work. Applying this model to the proton
``free'' cross sections are found  to be in good
agreement with the present and previously measured free cross-sections
and in reasonable agreement
with theoretical predictions based on multipole analyses
below 320 MeV. For the neutron no previous data are availble to
compare with. The ``free'' neutron cross sections obtained here 
also reasonably agree with theory up to 320 MeV. At higher energies there is
significant disagreement for both proton and neutron which is not 
presently understood.

\section{Acknowledgements}

{\acknowledgement
This work was supported by Deutsche
    Forschungsgemeinschaft (SFB\,201, SFB\,443, Schwerpunktprogramm
    1034 through contracts DFG-Wi1198 and DFG-Schu222), and by the
    German Russian exchange program 436 RUS 113/510.
  One of the authors (M.I.L.) highly appreciates the hospitality of the
  II.\@ Physi\-ka\-li\-sches Institut der  Universit\"at  G\"ottingen where
  part of the work was done. He is also very grateful to S. Kamalov
  for a computer code for the $\pi N$ partial amplitudes and to
  R.  Machleidt for a computer code for the CD-Bonn potential.  }
The authors also acknowledge the excellent support of the accelerator
group of MAMI.

\clearpage

%%%%%%%%%%%%%%%%%%%%%%%%%%%%%%%%%%%%%%%%%%%%%%%%%%%%%%%%%%%%%%%%%%%%%
%%%%%%%%%%%%%%%%%%%%%%%%%%%%%%%%%%%%%%%%%%%%%%%%%%%%%%%%%%%%%%%%%%%%%
%%%
%%%   T  A  B  L  E  S
%%%
%%%%%%%%%%%%%%%%%%%%%%%%%%%%%%%%%%%%%%%%%%%%%%%%%%%%%%%%%%%%%%%%%%%%%
%%%%%%%%%%%%%%%%%%%%%%%%%%%%%%%%%%%%%%%%%%%%%%%%%%%%%%%%%%%%%%%%%%%%%
\begin{appendix}

\begin{table}
  \caption{\label{table_qf_proton_pi0}The energy dependence of the
  triple differential cross-section of the reaction
           $d(\gamma,{\pi^0}p)n$ in CpQFP at
           $\theta^{\rm lab}_\pi=136.2^{\circ}$.  The statistical error is
           given. The systematic experimental error amounts to 4.4\%.}
\centering
  \begin{tabularx}{.8\columnwidth}{D{.}{.}{3}XD{.}{.}{2}@{~}c@{~}D{.}{.}{2}X}
    \hline\noalign{\smallskip}
    \multicolumn{1}{c}{$E_\gamma, \mathrm{MeV}$} &
    \multicolumn{5}{c}{$\left( \frac{d^3\sigma}{d\Omega_\pi
          d\Omega_p dE_p} \right)^\text{CpQFP} \left[
      \frac{\mu \text{b}}{\text{MeV sr}^2}\right] $} \\
    \noalign{\smallskip}\hline\noalign{\smallskip}
    254.3 && 3.61 & $\pm$ & 0.04 & \\
    273.5 && 5.42 & $\pm$ & 0.05 & \\
    292.7 && 7.88 & $\pm$ & 0.05 & \\
    312.0 && 9.67 & $\pm$ & 0.06 & \\
    331.2 && 9.78 & $\pm$ & 0.07 & \\
    350.4 && 8.66 & $\pm$ & 0.09 & \\
    369.4 && 6.88 & $\pm$ & 0.09 & \\
    388.5 && 5.27 & $\pm$ & 0.08 & \\
    \noalign{\smallskip}\hline
  \end{tabularx}
\end{table}

\begin{table}
  \caption{\label{table_qf_neutron_pi0}The energy dependence of the triple
           differential cross-section of the reaction
           $d(\gamma,{\pi^0}n)p$ in CnQFP at
           $\theta^{\rm lab}_\pi=136.2^{\circ}$. The statistical error is
           given. The systematic experimental error amounts to 9.0\%.}
           \centering
  \begin{tabularx}{.8\columnwidth}{D{.}{.}{3}XD{.}{.}{2}@{~}c@{~}D{.}{.}{2}X}
    \hline\noalign{\smallskip}
    \multicolumn{1}{c}{$E_\gamma, \mathrm{MeV}$} &
    \multicolumn{5}{c}{$\left( \frac{d^3\sigma}{d\Omega_\pi
          d\Omega_n dE_n} \right)^\text{CnQFP}\left[
       \frac{\mu \text{b}}{\text{MeV sr}^2}\right]$} \\
    \noalign{\smallskip}\hline\noalign{\smallskip}
    211.1 && 0.33 & $\pm$ & 0.02 & \\
    230.2 && 0.87 & $\pm$ & 0.03 & \\
    249.4 && 1.80 & $\pm$ & 0.04 & \\
    268.7 && 3.15 & $\pm$ & 0.06 & \\
    287.9 && 5.64 & $\pm$ & 0.08 & \\
    307.2 && 7.67 & $\pm$ & 0.10 & \\
    326.4 && 8.96 & $\pm$ & 0.12 & \\
    345.6 && 8.65 & $\pm$ & 0.15 & \\
    376.5 && 6.50 & $\pm$ & 0.11 & \\
    \noalign{\smallskip}\hline
  \end{tabularx}
\end{table}

\begin{table}
  \caption{\label{table_qffr_proton_pi0}The energy dependence of the
           differential cross-section of ``free'' proton $\pi^0$
           photoproduction extracted from quasi-free data at
           $\theta^{\rm lab}_\pi=136.2^{\circ}$. The statistical
           error is given. The systematic experimental error 
           amounts to 4.4\%. The error of the conversion of quasi-free
           cross sections to ``free'' cross sections amounts to 4\%.
            }
  \centering
  \begin{tabular}{cccr}
    \hline \\
$E_\gamma^f\left[ \mbox{MeV}\right]$ &$\frac { d\sigma }{
      d\Omega^{lab}_\pi }
\left[
\frac {\mu \text{b}}{\text{sr}}\right] $ & $\theta^{cm}_\pi\left[ 
\text{deg}\right]$ & $\frac {
d\sigma }{ d\Omega^{cm}_\pi }\left[ \frac {\mu \text{b}}{\text{sr}}\right]$
\\
    \noalign{\smallskip}
    \noalign{\smallskip}\hline\noalign{\smallskip}
    251.5 & 5.43  $\pm$  0.06 & 146.7 &  8.54  $\pm$  0.09 \\
    270.6 & 7.07  $\pm$  0.06 & 146.8 & 11.23  $\pm$  0.10 \\
    289.8 & 8.73  $\pm$  0.05 & 147.1 & 14.04  $\pm$  0.08 \\
    309.0 & 9.84  $\pm$  0.06 & 147.3 & 16.04  $\pm$  0.10 \\
    328.2 & 8.12  $\pm$  0.06 & 147.6 & 13.42  $\pm$  0.10 \\
    347.3 & 6.36  $\pm$  0.07 & 147.8 & 10.67  $\pm$  0.12 \\
    366.3 & 4.59  $\pm$  0.06 & 148.1 &  7.82  $\pm$  0.10 \\
    385.4 & 3.26  $\pm$  0.05 & 148.4 &  5.64  $\pm$  0.09 \\
    \noalign{\smallskip}\hline
  \end{tabular}
\end{table}

\begin{table}
  \caption{\label{table_qffr_neutron_pi0}The energy dependence of the
           differential cross-section of ``free'' neutron $\pi^0$
           photoproduction extracted from the quasi-free data at
           $\theta^{\rm lab}_\pi=136.2^{\circ}$. The statistical error is
           given. The systematic experimental error amounts to 9.0\%.
           The error of the conversion of quasi-free to ``free'' cross
           sections amounts to 4\%. 
           }
  \centering
  \begin{tabular}{cccr}
    \hline \\
$E_\gamma^f \left[ \mbox{MeV}\right]$ &$\frac { d\sigma }{ d\Omega^{lab}_\pi }
\left[\frac {\mu \text{b}}{\text{sr}}\right] $ & $\theta^{cm}_\pi
\left[\text{deg}\right]$ & $\frac {
d\sigma }{ d\Omega^{cm}_\pi } \left[ \frac {\mu \text{b}}{\text{sr}}\right]$
\\
    \noalign{\smallskip}
    \noalign{\smallskip}\hline\noalign{\smallskip}
    208.4 & 0.78  $\pm$  0.04 & 146.7 &   1.22 $\pm$ 0.08 \\
    227.4 & 1.68  $\pm$  0.05 & 146.5 &   2.62 $\pm$ 0.09 \\
    246.6 & 3.02  $\pm$  0.06 & 146.6 &   4.74 $\pm$ 0.11 \\
    265.8 & 4.53  $\pm$  0.08 & 146.8 &   7.17 $\pm$ 0.14 \\
    285.0 & 6.86  $\pm$  0.09 & 147.0 &  10.99 $\pm$ 0.16 \\
    304.3 & 8.11  $\pm$  0.10 & 147.2 &  13.17 $\pm$ 0.18 \\
    323.4 & 8.07  $\pm$  0.10 & 147.5 &  13.29 $\pm$ 0.18 \\
    342.6 & 6.80  $\pm$  0.12 & 147.7 &  11.36 $\pm$ 0.20 \\
    373.4 & 4.34  $\pm$  0.08 & 148.2 &   7.43 $\pm$ 0.12 \\
     \noalign{\smallskip}\hline
  \end{tabular}
\end{table}

\begin{table}
  \caption{\label{table_free_proton_pi0}The energy dependence of the
            differential cross-section of the reaction
           $p(\gamma,{\pi^0})p$ at $\theta^{\rm
            lab}_\pi=136.2^{\circ}$. 
The statistical
           error is given. The systematic experimental 
           error amounts to 4.4\%.}
 \centering
  \begin{tabular}{crcr}
    \hline \\
$E_\gamma\left[ \mbox{MeV}\right]$ &$\frac { d\sigma }{ d\Omega^{lab}_\pi }
\left[\frac {\mu \text{b}}{\text{sr}}\right]$ & $\theta^{cm}_\pi 
\left[ \text{deg}\right]$ & $\frac {
d\sigma }{ d\Omega^{cm}_\pi }\left[ \frac {\mu \text{b}}{\text{sr}}\right] $
\\ 
    \noalign{\smallskip}
    \noalign{\smallskip}\hline\noalign{\smallskip}
    254.3 & 5.32  $\pm$  0.08 & 146.7 &  8.38 $\pm$ 0.13 \\
    273.5 & 7.62  $\pm$  0.11 & 146.9 & 12.13 $\pm$ 0.18 \\
    292.7 & 9.79  $\pm$  0.10 & 147.1 & 15.77 $\pm$ 0.16 \\
    312.0 &11.06  $\pm$  0.12 & 147.3 & 18.07 $\pm$ 0.20 \\
    331.3 & 9.89  $\pm$  0.10 & 147.6 & 16.39 $\pm$ 0.17 \\
    350.4 & 7.91  $\pm$  0.09 & 147.9 & 13.31 $\pm$ 0.15 \\
    369.4 & 5.70  $\pm$  0.08 & 148.1 &  9.74 $\pm$ 0.14 \\
    388.5 & 4.08  $\pm$  0.07 & 148.4 &  7.08 $\pm$ 0.12 \\
    407.4 & 2.96  $\pm$  0.06 & 148.7 &  5.22 $\pm$ 0.11 \\
    427.3 & 2.10  $\pm$  0.05 & 149.0 &  3.76 $\pm$ 0.09 \\
    448.3 & 1.46  $\pm$  0.04 & 149.3 &  2.66 $\pm$ 0.07 \\
    469.0 & 1.11  $\pm$  0.04 & 149.5 &  2.06 $\pm$ 0.07 \\
  \noalign{\smallskip}\hline
  \end{tabular}
\end{table}

\end{appendix}

\clearpage

%%%%%%%%%%%%%%%%%%%%%%%%%%%%%%%%%%%%%%%%%%%%%%%%%%%%%%%%%%%%%%%%%%%%%
%%%%%%%%%%%%%%%%%%%%%%%%%%%%%%%%%%%%%%%%%%%%%%%%%%%%%%%%%%%%%%%%%%%%%
%%%
%%%   B  I  B  L  I  O  G  R  A  P  H  Y
%%%
%%%%%%%%%%%%%%%%%%%%%%%%%%%%%%%%%%%%%%%%%%%%%%%%%%%%%%%%%%%%%%%%%%%%%
%%%%%%%%%%%%%%%%%%%%%%%%%%%%%%%%%%%%%%%%%%%%%%%%%%%%%%%%%%%%%%%%%%%%%


\begin{thebibliography}{}

\bibitem{schmidt01}
 A. Schmidt {\em et al.}, Phys. Rev. Lett. {\bf 87}, 232501 (2001).

\bibitem{bernard96}
 V. Bernard, N. Kaiser, U.-G. Mei\ss ner, Z. Phys. C {\bf 70}, 483 (1996).

\bibitem{database}
 See the SAID database at the website {\tt http://gwdac.phys.gwu.edu}.

\bibitem{bacci72}
 C. Bacci {\em et al.}, Phys. Lett. B {\bf 39}, 559 (1972).

 \bibitem{hemmi73}
 Y. Hemmii {\em et al.}, Nucl. Phys.  B {\bf 55}, 333 (1973).

\bibitem{ando77}
 A. Ando {\em et al.}, Physik Daten, 1977 (unpublished).


\bibitem{LLP94}
M.I. Levchuk, A.I. L'vov, V.A. Petrun'kin, preprint FIAN
No. 86, 1986; Few-Body Syst. \textbf{16}, 101 (1994).

\bibitem{wissmann98}
F. Wissmann, M.I. Levchuk, M. Schumacher, Eur. Phys. J.
A \textbf{1}, 193 (1998).


\bibitem{kossert01}
K. Kossert {\em et al.}, Phys. Rev. Lett. \textbf{88}, 162301 (2002).

\bibitem{camen02}
M. Camen {\em et al.}, Phys. Rev. C \textbf{65}, 032202 (2002).

\bibitem{kossert03}
K. Kossert {\em et al.}, Eur. Phys. J A 16, 259 (2003).




%%%%%%%%%%%%%%%%% Section experiment

\bibitem{anthony91}
I. Anthony, J.D. Kellie, S.J. Hall, G.J. Miller, Nucl.
Instr. Meth. A \textbf{301}, 230 (1991),
S.J. Hall, G.J. Miller, R. Beck, P. Jennewein, Nucl. Instr. Meth.
A \textbf{368}, 698 (1996).

\bibitem{wiss94}F. Wissmann {\em et al.}, Phys. Lett. B \textbf{335}, 119
(1994).

\bibitem{wissmann99}
F. Wissmann {\em et al.}, Nucl. Phys. A \textbf{660}, 232 (1999).

\bibitem{seneca}G. v. Edel {\em et al.}, Nucl. Instr. Meth. A \textbf{365}, 224
  (1993).


\bibitem{brun93}
R. Brun {\em et al.}, GEANT Detector Description and Simulation Tool,
CERN Program Library Long Writeup W5013, Cern Geneva Switzerland (1994)\\
{\tt http://wwwinfo.cern.ch/asd/geant/index.html}.


% Theory of gd->pi^0np

\bibitem{laget78}
 J.M. Laget, Nucl. Phys. A {\bf 296}, 388 (1978).

\bibitem{laget81}
 J.M. Laget, Phys. Rep. {\bf 69}, 1 (1981).

\bibitem{LLP96}
 M.I. Levchuk, V.A. Petrun'kin, M. Schumacher,
 Z. Phys. A {\bf 355}, 317 (1996). %%% \pi^0 in the Delta region

\bibitem{LSW00}
 M.I. Levchuk, M. Schumacher, F. Wissmann,
 Nucl. Phys. A {\bf 675}, 621 (2000). %%% \pi^0 in the Delta region

\bibitem{LSW00_2}
 M.I. Levchuk, M. Schumacher, F. Wissmann, nucl-th/0011041.

\bibitem{schmidt96}
  R. Schmidt, H. Arenh\"ovel,  P. Wilhelm, Z. Phys. A {\bf 355},
  421 (1996).

\bibitem{darwish02}
 E.M. Darwish, H. Arenh\"ovel, M. Schwamb, Eur. Phys. J A {\bf 16}, 111 (2003).

\bibitem{krusche99}
  B. Krusche {\em et al.}, Eur.  Phys.  J.  A {\bf 6}, 309 (1999).

\bibitem{siodl01}
  U. Siodlaczek~\etal, Eur.  Phys.  J.  A {\bf 10}, 365 (2001) and
  U. Siodlaczek, PhD Thesis, University of T\"ubungen, 2000.

\bibitem{LSW03}
 M.I. Levchuk, M. Schumacher, F. Wissmann, in preparation.

%%%%%%%% SENECA calibration

%\bibitem{edel92}G. v. Edel, Diploma Thesis, Universit\"at G\"ottingen, 1992.

%\bibitem{gall93}G. Galler, Diploma Thesis, Universit\"at G\"ottingen,
%1993.

%\bibitem{maas95}R. Maa\ss, Diploma Thesis, Universit\"at G\"ottingen,
%1995.

%\bibitem{zeit96}
%C. Zeitnitz, T.A. Gabriel, "The GEANT-CALOR Interface User's Guide"
%Universit\"at Mainz, Oak Ridge National Laboratory, October (1996).

%\bibitem{fese85}
%H. Fesefeldt, "The simulation of hadronic showers, physics and
%applications", Technical Report PITHA 85-02, RWTH Aachen (1985).


%\bibitem{lara01}G. Galler {\em et al.}, Phys. Lett. B \textbf{501}, 245 (2001).

%\bibitem{wolf01} S. Wolf {\em et al.}, Eur. Phys. J. A {\bf 12}, 231 (2001).
%


\bibitem{arndt02}
R.A. Arndt, W.J. Briscoe, I.I. Strakovsky, R.L. Workman, Phys. Rev. C \textbf{66}, 055213
(2002) and the code SAID.

\bibitem{drechsel99}
D. Drechsel, O. Hanstein, S.S. Kamalov, L. Tiator,
Nucl. Phys. A \textbf{645}, 145 (1999) and the code MAID. 


%%%%%%%%%%%%%%%%%%% NN-potentials
\bibitem{OBEPR1} R. Machleidt, K. Holinde, Ch. Elster, Phys. Rep. {\bf 149},
  1 (1987).

\bibitem{OBEPR2} R. Machleidt, Adv. Nucl. Phys. {\bf 19}, 189 (1989).

\bibitem{CD-Bonn} R. Machleidt, Phys. Rev. C {\bf 63},  024001 (2001).

\bibitem{PEST1} J. Haidenbauer, W. Plessas, Phys. Rev. C {\bf 30},
1822 (1984).

\bibitem{PEST2} J. Haidenbauer, W. Plessas, Phys. Rev. C {\bf 32},
1424 (1985).

\bibitem{Paris} M. Lacombe {\em et al.}, Phys. Rev. D {\bf 12}, 1495 (1975).

%%%%%%%%%%%%%%%%%%% piN-scattering

\bibitem{hung94}
 C.T. Hung, S.N. Yang, T.-S.H. Lee, J. Phys. G {\bf 20}, 1531 (1994).

\bibitem{hung01}
 C.T. Hung, S.N. Yang, T.-S.H. Lee, Phys. Rev. C {\bf 64}, 034309 (2001).

\bibitem{kamalov99}
S.S. Kamalov, S.N. Yang, Phys. Rev. Lett. {\bf 83}, 4494 (1999). 

\bibitem{kamalov01}
S.S. Kamalov~\etal, Phys. Lett. B {\bf 522}, 27 (2001). 

% Experiments on gp->pi(0)p

\bibitem{obukh03}
I.T. Obukhovsky~\etal, nucl-th/0212110.

\bibitem{genzel74}
  H. Genzel {\em et al.}, Z. Phys. {\bf 268}, 43 (1974). 

\bibitem{fuchs96}
  M. Fuchs  {\em et al.}, Phys. Lett. B {\bf 368}, 20 (1996).

\bibitem{haerter96}
  F. H\"arter, PhD thesis, Mainz University, 1996.


\end{thebibliography}
\end{document}